\documentclass[apj]{emulateapj}

\submitted{ApJ, \textbf{696}, L129--L132 (2009) [Accepted 02/04/2009]}

\newcommand{\be}{\begin{equation}}
\newcommand{\ee}{\end{equation}}
\newcommand{\bea}{\begin{eqnarray}}
\newcommand{\eea}{\end{eqnarray}}
\newcommand{\ha}{HI}
\newcommand{\hm}{H$_2$}

\newcommand{\correction}{\zeta}
\newcommand{\h}{h}

\newcommand{\f}{R_{\rm mol}}
\newcommand{\fc}{\f^{\rm c}}
\newcommand{\fg}{\f^{\rm galaxy}}
\newcommand{\fu}{\f^{\rm cosmic}}
\newcommand{\rd}{r_{\rm disk}}

\newcommand{\rvir}{r_{\rm vir}}

\newcommand{\mass}{M}
\newcommand{\msun}{{\rm M}_{\odot}}
\newcommand{\lsun}{{\rm L}_{\odot}}

\newcommand{\mg}{\mass_{\rm g}}
\newcommand{\mha}{\mass_{\rm HI}}
\newcommand{\mhm}{\mass_{{\rm H}_2}}

\newcommand{\msdisk}{\mass_{\rm s}^{\rm disk}}

\newcommand{\Omegaha}{\Omega_{\rm HI}}
\newcommand{\Omegahm}{\Omega_{{\rm H}_2}}

\newcommand{\Omegahb}{\Omega_{\rm HI+H_2}}
\newcommand{\rhoc}{\rho_{\rm c}}

\newcommand{\rhoha}{\rho_{\rm HI}}
\newcommand{\rhohm}{\rho_{\rm H_2}}

\newcommand{\rhosfr}{\rho_{\rm SFR}}

\newcommand{\sfeha}{{\rm SFE_{HI}}}
\newcommand{\sfehm}{{\rm SFE_{H_2}}}

\shorttitle{Cosmic evolution of \ha~and \hm}
\shortauthors{Obreschkow et al.}

\begin{document}

\title{The Cosmic Decline in the \hm/\ha-Ratio in Galaxies}

\author{D. Obreschkow and S. Rawlings}
\affil{Astrophysics, Department of Physics, University of Oxford, Keble Road, Oxford, OX1 3RH, UK}

\begin{abstract}
We use a pressure-based model for splitting cold hydrogen into its atomic (\ha) and molecular (\hm) components to tackle the co-evolution of \ha, \hm, and star formation rates (SFR) in $\sim\!3\cdot10^7$ simulated galaxies in the Millennium Simulation. The main prediction is that galaxies contained similar amounts of \ha~at redshift $z\approx1-5$ than today, but substantially more \hm, in quantitative agreement with the strong molecular line emission already detected in a few high-redshift galaxies and approximately consistent with inferences from studies of the damped Lyman-$\alpha$ absorbers seen in the spectra of quasars. The cosmic \hm/\ha-ratio is predicted to evolve monotonically as $\Omegahm/\Omegaha\propto(1+z)^{1.6}$. This decline of the \hm/\ha-ratio as a function of cosmic time is driven by the growth of galactic disks and the progressive reduction of the mean cold gas pressure. Finally, a comparison between the evolutions of \ha, \hm, and SFRs reveals two distinct cosmic epochs of star formation: an early epoch ($z\gtrsim3$), driven by the evolution of $\Omegahb(z)$, and a late epoch ($z\lesssim3$), driven by the evolution of $\Omegahm(z)/\Omegaha(z)$.
\end{abstract}


\section{Introduction and key idea}\label{section_introduction}

Neutral hydrogen is the fuel for the formation of stars. The cosmic star formation rate (SFR) density as inferred from ultraviolet, far-infrared, and submillimeter observations increases by an order of magnitude from redshift $z=0$ to $z=2$ \citep{Hopkins2007}. Hence, neutral hydrogen in early galaxies was either more abundant or transformed into stars more efficiently than today.

A useful quantity in this context is the star formation efficiency (SFE) of a galaxy, defined as the SFR divided by the gas mass. The weak cosmic evolution of the density of neutral atomic hydrogen (\ha), derived from Lyman-alpha absorption against distant quasars \citep{Lah2007,Pontzen2009}, indicates a strongly increased SFE at high $z$. But recent detections of strong molecular line emission in ordinary galaxies at $z=1.5$ \citep{Daddi2008} suggest that the SFEs of these galaxies are similar to those seen today. The seeming contradiction between these two conclusions arises from the conceptual confusion of SFEs inferred from galactic \ha~with those inferred from \hm. In fact, it is crucial to distinguish between the two quantities $\sfeha\equiv{\rm SFR}/\mha$ and $\sfehm\equiv{\rm SFR}/\mhm$. In principle, there is no contradiction between the detected strong cosmic evolution of $\sfeha$ and the weak evolution of $\sfehm$ -- these empirical findings could simply imply that the \hm/\ha-mass ratios $\fg$ of galaxies increase substantially with $z$.

In this letter, we show that there is indeed strong theoretical support for such an increase of $\fg$ with $z$ in regular galaxies. This evolution is driven by the approximate scaling of galaxy sizes as $(1+z)^{-1}$ predicted by dark matter theory \citep{Gunn1972} and confirmed by observations in the Ultra Deep Field \citep{Bouwens2004}. Hence, the cold gas disks at high redshift must, on average, be denser than today. Combining this prediction with the relation between gas pressure and \hm/\ha-ratios in nearby galaxies \citep[e.g.][]{Blitz2006}, leads to the conclusion that $\fg$ must increase dramatically with $z$. Our quantitative predictions of this evolution rely on a recently presented semi-analytic numerical simulation of \ha~and \hm~in $\sim\!\!3\cdot10^7$ simulated galaxies \citep{Obreschkow2009b}, based on the Millennium Simulation \citep{Springel2005}.

Section \ref{section_simulation} overviews our simulation method and the model for the \hm/\ha-ratio in galaxies. In Section \ref{section_results}, we present and interpret the predicted evolution of galactic \ha~and \hm~and their relation to star formation. Section \ref{section_obs} compares these predictions to empirical data, and Section \ref{section_conclusion} summarizes our key conclusions.

\section{Simulating \ha~and \hm~in galaxies}\label{section_simulation}

\subsection{Physical model for galactic \hm/\ha-ratios}\label{subsection_model}

In virtually all regular galaxies in the local Universe, whether spirals \citep[e.g.][]{Leroy2008} or ellipticals \citep[e.g.][]{Young2002}, the cold gas resides in a flat disk. Some observations of CO at $z\approx2$ \citep{Tacconi2006} suggest that even at high redshift most cold gas lies in disks. Based on this evidence, we have recently introduced a model for the distributions of \ha~and \hm~in regular galaxies \citep{Obreschkow2009b}, assuming that all cold gas resides in a flat symmetric disk with an exponential surface density profile and that the local \hm/\ha-ratio is dictated by the kinematic gas pressure \citep{Blitz2006,Leroy2008}. Within these assumptions, we could show that the \hm/\ha-mass ratio $\fg$ of an entire galaxy is given by
\be\label{eqfg}
  \fg = \big(3.44\,{\fc}^{-0.506}+4.82\,{\fc}^{-1.054}\big)^{-1},
\ee
where $\fc$ represents the \hm/\ha-ratio at the galaxy center. $\fc$ can be approximated as
\be\label{eqfc}
  \fc =\!\left[11.3\,{\rm m^4 kg^{\!-2}}\rd^{-4}\mg\big(\mg\!+\!0.4\,\msdisk\big)\right]^{0.8},
\ee
where $\rd$ is the exponential scale radius of the disk, $\mg$ is the total cold gas mass, and $\msdisk$ is the stellar mass in the disk. Eqs.~(\ref{eqfg},\ref{eqfc}) constitute a physical model to estimate $\fg$ in regular galaxies based on $\msdisk$, $\mg$, and $\rd$. In order to predict the cosmic evolution of $\fg$, we therefore require a model for the co-evolution of $\msdisk$, $\mg$, and $\rd$ in galaxies. To this end, we adopted the virtual galaxy catalog of the Millennium Simulation described in Section \ref{subsection_simulation}. The limitations of the model of Eqs.~(\ref{eqfg},\ref{eqfc}) and their impact on the predicted \hm/\ha-ratios are discussed in \citep{Obreschkow2009b}.

\subsection{\ha~and \hm~in the Millennium Simulation}\label{subsection_simulation}

The Millennium Simulation \citep{Springel2005} is an $N$-body simulation within the $\Lambda$CDM cosmology of $\sim\!\!10^{10}$ gravitationally interacting particles in a periodic box of comoving volume $(500~\h^{-1}~\rm Mpc)^3$, where $H_0=100\,h\rm\,km\,s^{-1}\,Mpc^{-1}$ and $\h=0.73$. The evolving large-scale structure generated by this simulation served as the skeleton for the simulation of $\sim\!\!3\cdot10^7$ galaxies at the halo centers. In the ``semi-analytic'' approach adopted by \citet{DeLucia2007}, galaxies were considered as simplistic objects with a few global properties that are evolved stepwisely using a list of physical prescriptions. For example, the total amount of cold hydrogen (\ha+\hm) in a galaxy is defined by the history of the net accretion, which in the model consists of (i) the infall of gas from the hot halo, (ii) the loss of gas by star formation, and (iii) outflows driven by supernovae and active galactic nuclei. Star formation in each galaxy is tackled using a law, where all cold gas above a critical surface density is transformed into stars on a timescale proportional to the dynamical time of the disk \citep[for details see][]{Croton2006}.

In \citet{Obreschkow2009b}, we applied the model of Section \ref{subsection_model} to the simulated galaxies in the catalog of \citet{DeLucia2007} (``DeLucia-catalog''), to split their cold hydrogen masses into \ha~and \hm. Our simulation successfully reproduced many local observations of \ha~and \hm, such as mass functions (MFs), mass--diameter relations, and mass--velocity relations. Yet, the high-redshift predictions are inevitably limited by the semi-analytic recipes of the DeLucia-catalog. The most uncertain recipes are those related to mergers (e.g.~feedback of black hole coalescence and starbursts), but they have a minor effect on the cosmic space densities of \ha~and \hm, since most cold gas in the simulation is found in regular disk galaxies\footnote{By contrast, a significant fraction of the \emph{stars} at $z=0$ is in massive elliptical galaxies with violent merger histories, but even those galaxies formed most stars in their spiral progenitors.} with at most minor merger histories. However, inaccurate prescriptions for isolated galaxies could significantly affect the space densities of \ha~and \hm, and it may well become necessary to refine our simulation as improved semi-analytic methods come on line.

\section{Results}\label{section_results}

\subsection{Predicted evolution of \ha~and \hm}\label{subsection_evolution}

\begin{figure}[h]
  \includegraphics[width=\columnwidth]{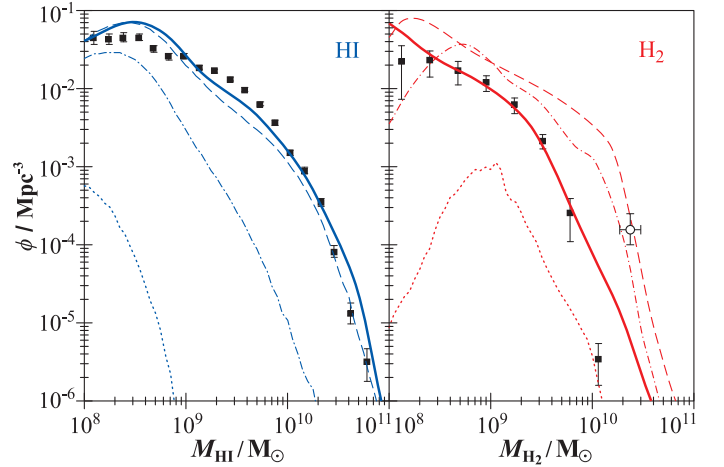}
  \caption{\small{MFs of \ha~and \hm. Lines show the simulation results at $z=0$ (solid), $z=2$ (dashed), $z=5$ (dash-dotted), $z=10$ (dotted). Square dots represent the empirical data and 1-$\sigma$ scatter at $z=0$ \citep{Zwaan2005,Obreschkow2009a}, and the open circle represents our density estimate at $z=1.5$ (Section \ref{section_obs}) based on \cite{Daddi2008}.}}
  \label{fig_mfs}
\end{figure}

Fig.~\ref{fig_mfs} shows the predicted evolution of the \ha-MF and \hm-MF, i.e.~the comoving space densities of sources per logarithmic mass interval. The predictions at $z=0$ roughly agree with available observational data, but the obvious differences, such as the spurious bumps around $\mha\approx10^{8.5}$ and $\mhm\approx10^8$ (a mass resolution limit), have been discussed in \cite{Obreschkow2009b}.

\begin{figure}[h]
  \includegraphics[width=\columnwidth]{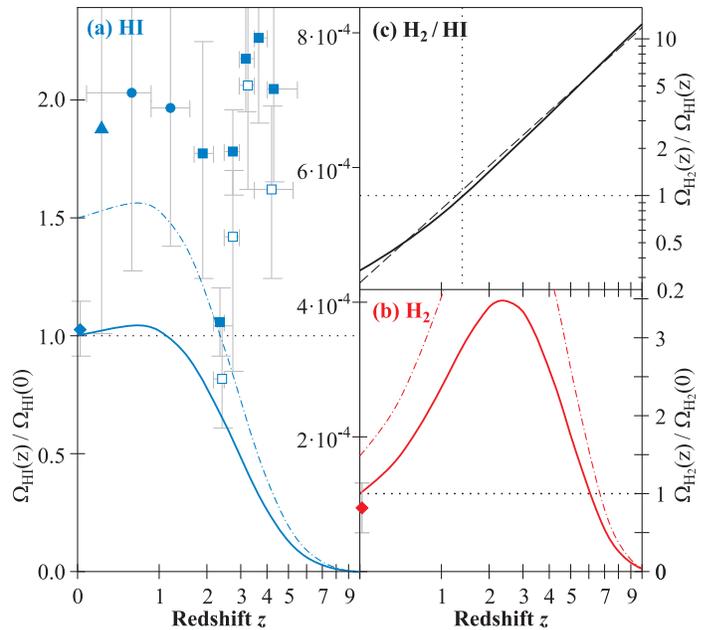}
  \caption{\small{Cosmic evolution of the fractional space densities of \ha~and \hm. Solid lines represent the simulated evolution of $\Omegaha$ (a), $\Omegahm$ (b), and $\Omegahm/\Omegaha$ (c). The dashed line in panel (c) is the power-law fit for $\Omegahm/\Omegaha$ given in Eq.~(\ref{eqfu}). The points represent the observations described in Section \ref{section_obs}. The dash-dotted lines in panels (a) and (b) represent the evolution of $\Omegaha$ and $\Omegahm$, if the correction factor $\correction$ in \cite{Obreschkow2009b} is set to $\correction=1$, i.e.~the total cold gas mass of the DeLucia-catalog is not corrected. We originally introduced this correction to fit the sum of the local space densities of \ha~and \hm~(shown as diamonds).}}
  \label{fig_omegah}
\end{figure}

The predicted \ha-masses remain roughly constant from $z=0$ to $z=2$, while \hm-masses increase dramatically. These different evolutions are also reflected in the comoving space densities $\Omegaha\equiv\rhoha/\rhoc$ and $\Omegahm\equiv\rhohm/\rhoc$, where $\rhoc(z)=3H^2(z)/(8\pi G)$ is the critical density for closure. Here, $\Omegaha$ and $\Omegahm$ only account for gas in galaxies, excluding unbound \ha~between the first galaxies \citep{Becker2001} or possible \hm~in haloes \citep{Pfenniger1994}. The simulated functions $\Omegaha(z)$ and $\Omegahm(z)$ are shown in Figs.~\ref{fig_omegah}a, b, while Fig.~\ref{fig_omegah}c represents their ratio $\fu(z)\equiv\Omegahm(z)/\Omegaha(z)$, which is closely described by the power-law
\be\label{eqfu}
  \fu(z)\approx0.3\cdot(1+z)^{1.6}.
\ee
The simulation yields $\fu(0)\approx0.3$ and finds the crossover, $\fu(z)=1$, at $z\approx1.4$. Our model predicts that Eq.~(\ref{eqfu}) extends to epochs, where the first galaxies formed, but this prediction is likely to breakdown at the highest redshifts, where the formation of \hm~was inhibited by the lack of metals \citep{Abel2000}.

Physically, the strong evolution of \hm/\ha~is essentially driven by the size-evolution of galaxies and their haloes. The Millennium Simulation assumes that the virial radius $\rvir$ of a spherical halo always encloses a mass with an average density 200-times above the critical density $\rhoc\propto H^2$ \citep{Croton2006}. Hence, for a fixed halo mass, $\rvir\propto H^{-2/3}$. In a flat Universe this implies
\be\label{eqproprvir}
  \rvir\propto\big[\Omega_{\rm m}(1+z)^3+\Omega_\Lambda\big]^{-1/3},
\ee
which asymptotically tends to $\rvir\propto(1+z)^{-1}$ for high $z$. By virtue of the theory of \citet{Fall1980}, this cosmic scaling of $\rvir$ results in a similar scaling of the disk radius, i.e.~$\rd\propto(1+z)^{-1}$, consistent with observations in the Ultra Deep Field \citep{Bouwens2004}.

For the gas-dominated galaxies in the early Universe, Eq.~(\ref{eqfc}) reduces to $\fc\propto\rd^{-3.2}\mg^{1.6}$. Yet, the cold gas masses $\mg$ of individual galaxies in the simulation evolve weakly with cosmic time, due to a self-regulated equilibrium between the net inflow of gas and star formation. In fact, most of the  evolution of $\Omegahb$ in the redshift range $z\approx3-10$ is due to the build-up of new galaxies. Therefore, $\fc\propto\rd^{-3.2}\propto(1+z)^{3.2}$. At redshifts $z\approx1-10$, $\fc$ typically takes values between $10$ and $10^4$, such that Eq.~(\ref{eqfg}) can be approximated as $\fg\propto\,{\fc}^{0.5}$. Hence, $\fg\propto(1+z)^{1.6}$, which explains the scaling of Eq.~(\ref{eqfu}).

The cosmic evolution of $\Omegahm$ shown in Fig.~\ref{fig_omegah} can be divided in two epochs: The \emph{early epoch} ($z\gtrsim3$), where $\Omegahm$ increases with cosmic time, and the \emph{late epoch} ($z\lesssim3$), where $\Omegahm$ decreases with time. In the early epoch, $\fg>1$ implies $\Omegahm\approx\Omegahb$, and hence the growth of $\Omegahm$ reflects the general increase of $\Omegahb$ due to the intense assembly of new galaxies. In the late epoch, $\fg\lesssim1$ implies that $\Omegahm\approx\fu\Omegahb$. At this epoch the formation of the massive galaxies in the simulation is completed, i.e.~$\Omegahb(z)\approx{\rm const}$ and $\Omegahm\propto\fu$. Thus the decrease of $\Omegahm$ in this late epoch is driven by cosmic decline in $\fu$ or, physically, by the cosmic evolution of pressure.

\subsection{Link between \ha, \hm, and star formation}\label{subsection_sf}

To discuss the global cosmic evolution of the efficiencies $\sfeha$ and $\sfehm$ (Section \ref{section_introduction}), we shall define
\be\label{eqglobalsfedef}
  \langle\sfeha\rangle\equiv\rhosfr/\rhoha\,,\quad\langle\sfehm\rangle\equiv\rhosfr/\rhohm,
\ee
where $\rhoha\propto\Omegaha$, $\rhohm\propto\Omegahm$, and $\rhosfr$ denote the comoving space densities of \ha, \hm, and SFR.

In the semi-analytic recipes of the DeLucia-catalog, SFRs are estimated from the gas density and the dynamical time scale of the disk (Section \ref{subsection_simulation}). This Schmidt--Kennicutt law \citep{Schmidt1959,Kennicutt1998} for star formation makes similar predictions to models based on cold gas pressure\citep[e.g.][]{Blitz2006}, and therefore the SFRs in the DeLucia-catalog are, by default, approximately consistent with our model to split cold hydrogen into \ha~and \hm. The evolutions of $\langle\sfeha\rangle$ and $\langle\sfehm\rangle$ predicted by the simulation again reflect the marked difference between \ha~and \hm. They are approximated ($\sim\!\!20\%$ relative error) by the power-laws,
\bea
  \langle\sfeha\rangle/[Gyr^{-1}] & = & 0.23\,(1+z)^{2.2}, \label{eqsefha} \\
  \langle\sfehm\rangle/[Gyr^{-1}] & = & 0.75\,(1+z)^{0.6}, \label{eqsefhm}
\eea
out to $z\approx8$.

\begin{figure}[t]
  \includegraphics[width=\columnwidth]{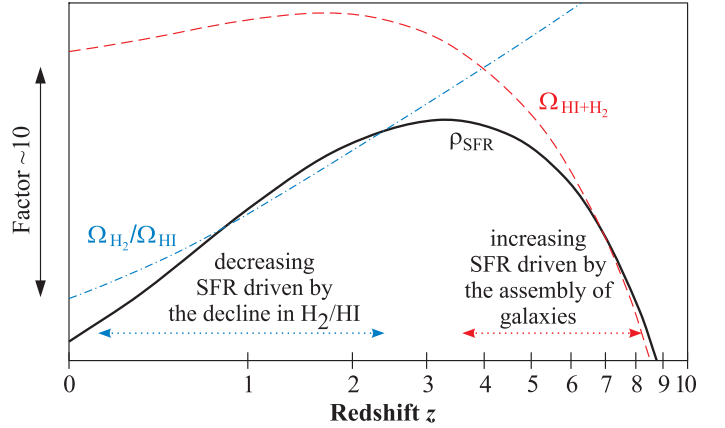}
  \caption{A simplistic model for the cosmic history of star formation.}
  \label{fig_simple_model}
\end{figure}

Due to the low power in Eq.~(\ref{eqsefhm}) $\rhosfr(z)$ is approximately proportional to $\Omegahm(z)$. We can therefore apply the two cosmic epochs of $\Omegahm(z)$ introduced in Section \ref{subsection_evolution} to the history of star formation (see Fig.~\ref{fig_simple_model}): In the \emph{early epoch} ($z\gtrsim3$), $\rhosfr$ increases with cosmic time, proportionally to $\Omegahb$. This increase traces the dramatic assembly of new galaxies. In the \emph{late epoch} ($z\lesssim3$), $\rhosfr$ decreases roughly proportionally to $\Omegahm/\Omegaha$. This epoch is driven by the cosmic evolution of pressure (or density) in galactic disks. This interpretation of the history of star formation does not, in fact, conflict with the picture that star formation is ultimately defined by the accreted cold gas mass (see Section \ref{subsection_simulation}) and a Schmidt--Kennicutt law for transforming this gas into stars. Our \hm/\ha-based interpretation simply adds another layer to the causal chain, by suggesting that cold gas mass and density ultimately dictate the amount of molecular material available for star formation.

%%%% not in APJ Version
%%%%

The simulation also includes star formation via merger-driven starbursts, associated with the creation of the stellar spheroids of early-type spiral or elliptical galaxies. However, the cosmic star formation density caused by mergers only accounts for about $1\%$ of $\rhosfr$ in the semi-analytic simulation of the DeLucia-catalog. (This fraction should not be confused with the fraction of gas-rich or ``wet'' mergers, since only some of the cold gas involved in these mergers is efficiently turned into stars in the form of a starburst.)

\section{Comparison with observations}\label{section_obs}

The DeLucia-catalog and our post-processing to assign \ha~and \hm, rely on established data of the local Universe. Our simulated \ha- and \hm-properties at $z=0$ are consistent with all available observations, i.e.~MFs (see Fig.~\ref{fig_mfs}), disk sizes, and velocity profiles \citep{Obreschkow2009b}. In particular, the simulated values $\Omegaha(0)=3.5\cdot10^{-4}$ and $\Omegahm(0)=1.2\cdot10^{-4}$ are consistent with the values (diamonds in Fig.~\ref{fig_omegah}) derived from the MFs observed in \ha-~and CO-emission at $z\approx0$ \citep{Zwaan2005,Obreschkow2009a}. At $z>0$, the currently available data are sparse, especially in emission.

The only measurement of $\Omegaha$ in emission at intermediate redshift is based on the stacking of 121 galaxies at $z=0.24$ (\citealp{Lah2007}, triangle in Fig.~\ref{fig_omegah}). The detection is speculative (see Fig.~7 in \citealp{Lah2007}), but roughly consistent with our simulation. All other measurements of $\Omegaha$ at $z>0$ rely on absorption detections of damped Lyman-$\alpha$ systems (DLAs). Respective data points from \citet{Rao2006} (circles in Fig.~\ref{fig_omegah}) and \cite{Prochaska2005} (filled squares) are, taken together, inconsistent with the predicted values of $\Omegaha$. By contrast, \cite{Zwaan2005b} demonstrated that the population of \ha-galaxies in the local Universe can fully explain the column density distributions of DLAs out to $z=1.5$, consistent with the nearly absent evolution of $\Omegaha$ from $z=0$ to $z=1.5$ predicted by our simulation. At present it is therefore difficult to judge, whether the simulation is inconsistent with empirical data at these low redshifts covering $2/3$ of the age of the Universe. At higher redshifts, however, the measurements of $\Omegaha$ seem not reconcilable with the simulated result, and even accounting for gravitational lensing by the DLAs only corrects the empirical values of $\Omegaha$ by about $30\%$ (open squares in Fig.~\ref{fig_omegah}, \citealp{Prochaska2005}). The simulated values of $\Omegaha$ are likely to underestimate the real values by about a factor 2 -- a plausible offset given the long list of simplifying approximations required from the $N$-body Millennium Simulation to our final post-processing of hydrogen in galaxies. Much progress could be expected from treating \ha-masses and \hm-masses as separate quantities directly in the semi-analytic galaxy simulation. This would allow, for example, to refine the feedback-mechanisms for suppression of gas infall \citep[explained in][]{Croton2006}, such that \ha~can still be accreted, while the formation of \hm~and stars is inhibited. Such a semi-analytic setting would also allow the implementation of a recipe for the large-scale dissociation of molecular gas by the radiation of newly formed stars \citep{Allen1986}. Both examples would effectively increase the amount of \ha~in high-redshift galaxies.

The most representative high-redshift observations of molecular gas to-date rely on two \emph{normal} galaxies (\textit{BzK}-4171 and \textit{BzK}-21000) at $z\approx1.5$, reliably detected in CO(2--1) emission by \citet{Daddi2008}. Unlike other CO-sources at similar or higher $z$, these objects are ordinary massive galaxies with FIR-luminosities of $L_{\rm FIR}\approx10^{12}~\lsun$, selected only due to the availability of precise spectroscopic redshifts. From these two detections, we estimated the \hm-space density (empty circle in Fig.~\ref{fig_mfs}) as follows: The mass interval spans between the masses $\mhm\approx2\cdot10^{10}\msun$ and $\mhm\approx3\cdot10^{10}$, respectively obtained for \textit{BzK}-4171 and \textit{BzK}-21000 by applying the CO-to-\hm~conversion of $\alpha=1~\msun(\rm K~km~s^{-1}\,pc^{-2})^{-1}$ \citep{Daddi2008}. The space density of these CO-sources was approximated as the space density of FIR-sources at $L_{\rm FIR}\approx10^{12}~\lsun$, based on the fact that all (both) targeted galaxies with $L_{\rm FIR}\approx10^{12}~\lsun$ revealed similar CO-luminosities $L_{\rm CO}$. We estimate their space density to be $1-2\cdot10^{-4}~{\rm Mpc}^{-3}$ per unit of $\log(L_{\rm FIR})$ by extrapolating the FIR-luminosity functions (LFs) of \citet{Huynh2007}. Since $L_{\rm FIR}\propto L_{\rm CO}\propto\mhm$, we find roughly the same space density per unit $\log(\mhm)$. These result is consistent with the simulated \hm-MF at $z=2$ (Fig.~\ref{fig_mfs}).

Considering \hm-absorption studies, \citet{Curran2004} and \citet{Noterdaeme2008} have determined \hm/\ha-ratios in DLAs that showed \hm-absorption. They found \hm/\ha-ratios of $\sim\!\!10^{-6}$ to $\sim\!\!10^{-2}$ at $z\approx2-3$, clearly much smaller than our prediction for $\Omegahm/\Omegaha$. We argue that measurements of \hm/\ha~in DLAs do not trace $\Omegahm/\Omegaha$ since DLAs are by definition \ha-selected objects and \hm~has a much smaller space coverage than \ha. In fact, \hm-disks in galaxies are much smaller than \ha-disks, especially at high $z$ \citep{Obreschkow2009d}, and even inside the \hm-disks the coverage of \hm~is small compared to \ha~\citep[e.g.][]{Ferriere2001}. A more detailed explanation of why \hm-searches in DLAs are expected to be difficult was given by \citet{Zwaan2006} based on the analysis of CO-emission maps of local galaxies.

\section{Conclusions}\label{section_conclusion}

In this letter, we have predicted the cosmic evolution of \ha- and \hm-masses in $\sim\!3\cdot10^7$ simulated galaxies based on the Millennium Simulation. The predicted cosmic decline in the \hm/\ha-ratio is consistent with the weak cosmic evolution of $\Omegaha$ inferred from DLA-studies and recent observations revealing a significantly enhanced space density of \hm~at $z=1.5$ \citep{Daddi2008}.

Perhaps the most important conclusion is that \ha- and \hm-masses evolve very differently with cosmic time and therefore cannot be used as proportional tracers of one another, especially not for the purpose of high-redshift predictions. There is no contradiction between the large \hm-masses detected at high $z$, which imply values of $\sfehm$ similar to those in the local Universe, and the weak evolution of \ha, implying massively increased values of $\sfeha$ at high $z$.

\acknowledgments
This work is supported by the European Community Framework Programme 6, Square Kilometre Array Design Studies (SKADS), contract no 011938. The Millennium Simulation databases and the web application providing online access to them were constructed as part of the German Astrophysical Virtual Observatory. We also thank the anonymous referee for the helpful suggestions.

%\bibliography{../Bibliography/astro}
%\bibliographystyle{../Bibliography/my_mn2e}

\end{document}